\documentclass[aps,prl,final,twocolumn,letterpaper]{revtex4}
\usepackage{graphicx}
\usepackage{epstopdf}
\usepackage{amsmath}
\usepackage{physics}
\usepackage{bm} 
\usepackage{amssymb}
\usepackage{xcolor}
\usepackage{array}
\usepackage{multirow}
\usepackage{lipsum}
\usepackage[T1]{fontenc}
\usepackage{multirow}
\usepackage[english]{babel}
\usepackage{hyperref}

\begin{document}

\title{Strong-coupling phases of trions and excitons in electron-hole bilayers at commensurate densities}
\author{David~D.~Dai}
\author{Liang~Fu}
\affiliation{Department of Physics, Massachusetts Institute of Technology, Cambridge MA 02139, USA}

\date{\today}

\begin{abstract}
We introduce density imbalanced electron-hole bilayers at a commensurate 2 : 1 density ratio as a platform for realizing novel phases of electrons, excitons, and trions. Through the independently tunable carrier densities and interlayer spacing, competition between kinetic energy, intralayer repulsion, and interlayer attraction yields a rich phase diagram. By a combination of theoretical analysis and  numerical calculation, we find a variety of strong-coupling phases in different parameter regions, including quantum crystals of electrons, excitons, and trions. We also propose an ``electron-exciton supersolid'' phase that features electron crystallization and exciton superfluidity simultaneously. The material realization and experimental signature of these phases are discussed in the context of semiconductor transition metal dichalcogenide bilayers. 
\end{abstract}

\maketitle

\pagestyle{myheadings}
\thispagestyle{empty}

\maketitle

\textit{Introduction---} Recently, semiconductor transition metal dichalcogenide (TMD) heterostructures have emerged as an ideal platform for exploring quantum phases of matter. An extraordinarily rich variety of quantum states have been predicted and observed, including Mott-Hubbard and charge-transfer insulators \cite{mott1, mott2, mott3, mott4, mott5, ladder}, Wigner crystals \cite{mott3, wigner1, wigner2, wigner3, wigner4, wigner5, wigner6, wigner7, wigner8}, itinerant ferromagnets \cite{itinerant1, pseudogap1, itinerant2, itinerant3, itinerant4, itinerant5}, interfacial ferroelectrics \cite{ferroelectricity}, heavy Fermi liquids \cite{heavy1, heavy2, heavy3}, spin-polaron liquids \cite{pseudogap1, pseudogap2, pseudogap3}, as well as quantum spin Hall states \cite{anomalous1, spinHall} and quantum anomalous Hall states \cite{anomalous2, anomalous3, anomalous4, anomalous5, anomalous6, anomalous7, anomalous8, anomalous9, anomalous10, anomalous11}. Remarkably, all these electronic phases realized in a single material system are rooted in one common ground---the two-dimensional electron (or hole) gas in monolayer TMDs. Here, the large effective mass and reduced screening favors strong interactions, and moir\'e bands in TMD heterostructures further enrich the physics, leading to much of the observed phenomena. 

In addition to moir\'e physics, TMD heterostructures provide a material realization of electron-hole ($e$-$h$) bilayers featuring electrons and holes on spatially separated layers, whose densities can be independently tuned by top and bottom gate voltages \cite{ZengBilayer}. The coexistence of positively and negatively charged particles supports both repulsion between like charges and attraction between opposite charges, enabling new phases of matter. When the densities of electrons and holes are equal, interlayer excitons with intrinsic out-of-plane dipole moments form, which may support high-temperature exciton superfluidity in an electrical insulator. Recently, thermodynamic evidence of excitonic insulator ground states has been observed in TMD bilayer WSe$_2$/MoSe$_2$  with WSe$_2$ as the hole layer and MoSe$_2$ as the electron layer separated by insulating hBN layers that suppress recombination \cite{exciton1, exciton2}. Moreover, density imbalanced electron-hole bilayers have attracted increasing interest owing to the interaction between charge carriers and dipolar excitons \cite{inbalanced}.

In this work, we study the strong-coupling phases of an imbalanced electron-hole bilayer at commensurate electron and hole densities $n_e/n_h=2$, motivated by the following considerations. Because there is a net charge density $n=n_e-n_h\neq 0$, the Coulomb interaction between charged particles dominates in the low-density regime, favoring strong-coupling phases with crystalline order. This should be contrasted to the balanced electron-hole bilayer, where charge-neutral excitons condense into a superfluid at low densities because their mutual dipole-dipole interaction is parametrically weaker than the quantum kinetic energy. The particular choice of $n_e/n_h=2$ is also motivated by the prospect of three-body bound states known as trions, which are charge-$e$ composite particles made of two electrons in the same layer bound to a hole in the other layer \cite{TEPL_exciton_trion, dynamics_exciton_trion}. 

Through both theoretical analysis and numerical calculation, we find a number of ordered phases driven by strong interactions. These include bilayer electron-hole Wigner crystals, ``composite crystals'' of coexisting electrons and excitons, as well as exotic quantum phases without classical counterparts. In particular, we find a quantum supersolid made of electrons and excitons as well as a quantum crystal of trions in the low-density regime. Experimental signatures of these predicted phases in TMD bilayers are also discussed in the context of optical spectroscopy and transport measurements.

Before presenting our main results, we first consider the three-body problem of two electrons and one hole, which reside on spatially separated layers ($z=d$ and $0$) and mutually interact through Coulomb forces. Classically, the minimum energy configuration is an electron and a dipole that are far apart due to their residual repulsion. This dipole (equivalent to a classical exciton) consists of an electron and a hole sitting directly on top of each other, separated only by the layer distance $d$. For comparison, consider a ``trion'' charge cluster with the two electrons at $(\pm {\mathbf r}, d)$ and the hole at $(\mathbf{0},0)$. When $r=1 / \sqrt{2^{4/3} - 1} \approx 0.811 d$, the net force acting on each particle is zero. However, this force-balanced configuration has energy $-0.937/d$, which is higher than the energy $-1/d$ of a dipole plus an electron far away. Additionally, this classical ``trion'' has an unstable normal mode, where the hole moves towards one of the electrons along the $\mathbf{r}$ direction, making the classical ``trion'' a saddle point of the energy instead of a local minima.

Our above analysis of the classical three-body problem demonstrates that quantum mechanics is crucial for the formation of a trion bound state. As shown by numerical studies, the trion is the ground state of two electrons and one hole in the bilayer when $a_B/d >0.065$ assuming equal electron and hole masses \cite{Witham2018}, where $a_B = \frac{4\pi\epsilon\hbar^2}{e^2 m}$ is the Bohr radius (which vanishes in the classical limit $\hbar \rightarrow 0$ or $m\rightarrow \infty$). We can understand the trion's quantum origin heuristically. Start with the electron and the dipolar exciton separated by a large distance $r\gg d$, where they experience a static $1/r - 1 / \sqrt{r^2 + d^2} \propto d^2/r^3$ repulsion. On the other hand, the electron's in-plane electric field $\propto 1/r^2$ polarizes the exciton and lowers its energy through the second-order Stark effect by an amount $\delta E_s \propto - a_B^3 /r^4 $ to zeroth order in $d$. This quantum attraction dominates the classical electron-exciton repulsion over a range of distances below a crossover length  $r_c\sim a_B^3/d^2$, supporting a trion bound state for sufficiently small $d$. 

At finite charge density, an additional length scale appears: the average inter-particle distance $a\equiv 1/\sqrt{\pi n}$, with $n = n_h$ for $n_e/n_h=2$. With three length scales---the inter-particle distance $a$, the layer distance $d$, and Bohr radius $a_B$, competition between intralayer repulsion, interlayer electron-hole attraction, and quantum kinetic energy yields a rich phase diagram for the electron-hole bilayer, which we explore below by a combination of analytical and numerical methods.   

The Hamiltonian for the bilayer assuming equal electron and hole effective masses $m$ and $1/r$ Coulomb interactions is:
\begin{equation}\label{eq:model_hamiltonian}
\begin{split}
H &= \sum_{a=e,h} \sum_{s=\uparrow, \downarrow} \int\text{d}^2{\mathbf r} \bigg[ \psi^{a\dagger}_{s}({\mathbf r}) \left(\frac{-\nabla^2}{2} \right) \psi^a_{s}({\mathbf r}) \bigg]
\\ &+ \frac{1}{2} \int\text{d}^2{\mathbf r} \int\text{d}^2{\mathbf r'} \bigg[
\sum_{a=e,h} \frac{n^a({\mathbf r}) n^a({\mathbf r'})}{|\mathbf{r-r'}|} \bigg]
\\ &- \int\text{d}^2{\mathbf r} \int\text{d}^2{\mathbf r'} \bigg [\frac{n^e({\mathbf r}) n^h({\mathbf r'})}{\sqrt{|\mathbf{r-r'}|^2 + d^2}} \bigg]
\end{split}
\end{equation}
where $\psi^{e,h}({\mathbf r})$ denotes the electron or hole operator, $n^{a}({\mathbf r}) =\sum_s \psi^{a\dagger}_s({\mathbf r}) \psi^a_s({\mathbf r})$ is the density operator, and $s$ denotes the spin. We have already divided all quantities by their appropriate atomic units, i.e. lengths by the Bohr radius $a_B = \frac{4\pi\epsilon\hbar^2}{m e^2}$ (we use this definition throughout) and energies by the Hartree energy $E_h = \frac{\hbar^2}{ma_B^2}$. Unless stated otherwise, we assume for simplicity that the electron and hole masses are equal, noting that our main findings are qualitatively correct for a range of mass ratios.

\begin{figure}[t!]
  \centering
  \includegraphics[width=0.5\textwidth]{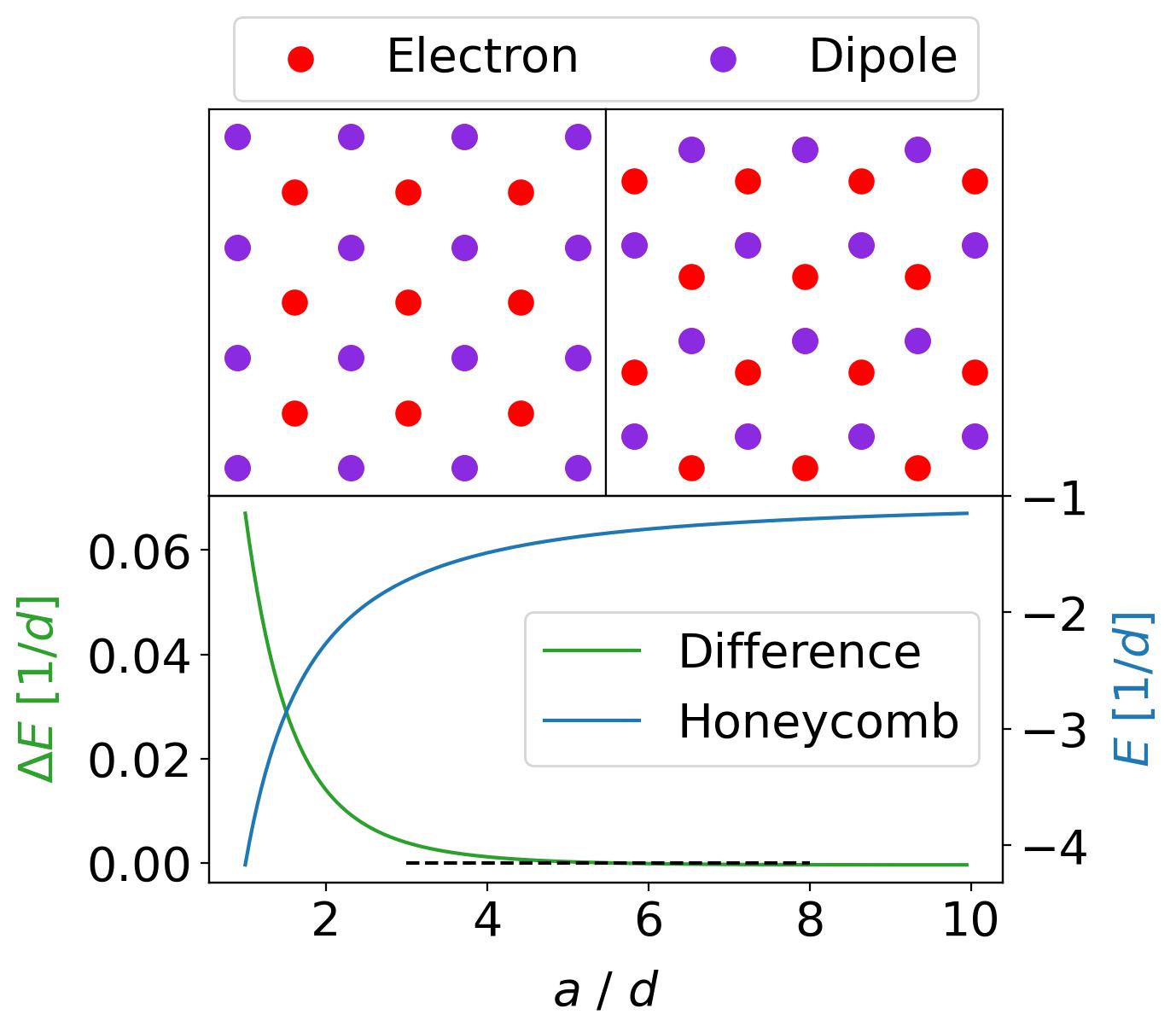}
  \caption{The classical phase diagram of the electron-hole bilayer at $n_e/n_h = 2$ for $a / d > 1$. The top panel depicts the checkerboard composite crystal (left) and the honeycomb composite crystal (right). The bottom panel shows the energy and energy difference of the two composite crystals, with a dashed line indicating the phase transition.}
  \label{fig:classical}
\end{figure} 

\textit{Classical Regime---} We first consider the classical limit defined by taking $a_B\rightarrow 0$ while keeping $d$ and $a$ fixed, or equivalently, $a/a_B\rightarrow \infty$ and $d/a_B \rightarrow \infty$ for a fixed $d/a$. In the classical limit, the phase diagram is characterized solely by the dimensionless ratio $d/a$. Both limits $d/a \gg 1$ and $d/a \ll 1$ can be understood analytically. When $d/a \gg 1$, the interlayer coupling is negligible and a triangular lattice Wigner crystal is formed independently in each layer. In the opposite regime $d/a \ll 1$, every hole pairs with an electron at the shortest possible distance $d$ to form a small out-of-plane dipole, leaving an equal number of excess electrons. These dipoles have a weak repulsion $1/r - 1 / \sqrt{r^2 + d^2} \propto d^2/r^3$ with the electrons, while electrons interact with each other through a strong Coulomb repulsion $1 /r$. Therefore, the electrons arrange themselves essentially independently of the dipoles to form a triangular lattice Wigner crystal. Once the electrons crystallize, the dipoles crystallize in half of the voids between the electrons to minimize the residual repulsion. The result is a ``composite crystal'' of electrons and dipoles, where electrons occupy one sublattice of a honeycomb and dipoles occupy the other.

To determine the classical ground state as a function of the charge density, we minimized the classical electrostatic energy. To handle the Coulomb interaction's long-ranged tail, we used Ewald summation \cite{supplement}. For the experimentally relevant parameter range $a/d>1$, we find two distinct composite crystals. For the low-density limit $a / d > 5.42$, the numerical results confirm our expectation of a honeycomb composite crystal. Interestingly, for $a / d < 5.42$ we find a ``checkerboard'' composite crystal consisting of two interpenetrating square lattices of electrons and dipoles. This is likely due to the increased electron-dipole repulsion at smaller $a / d$. We did not observe the formation of classical trions for any $a/d$, which is consistent with their instability. Once these two composite crystals were identified, we directly calculated their electrostatic energy as a function of $a / d$ and determined the phase transition point, which is shown in Fig. \ref{fig:classical}.

We now analyze the effect of quantum fluctuations around the classical composite crystals. At sufficiently small $a_B$, the leading quantum effect is the zero-point-motion of charges in the classical ground state. The root-mean-square displacement of charges increases with $a_B$, and when it becomes comparable to the lattice constant $a$, quantum melting of the crystal occurs (Lindemann criterion) \cite{LindemannCriterion}. Since our composite crystals have three charges per unit cell, there are a total of six phonon modes, including both acoustic and optical branches. Notably, the optical phonons are associated with the relative vibration of charges within the unit cell, which is absent in the canonical electron Wigner crystal.

When the layer distance $d$ is small compared to the inter-particle distance $a$, the electron-dipole repulsion in the composite crystal is much weaker than the electron-electron repulsion. Then we expect that the dipole's center of mass has the largest zero-point displacement, denoted as $\xi_d$. This is indeed confirmed by our direct calculation of the optical phonon frequencies at zero wavevector. Two types of optical phonons are present: the low-frequency one corresponds to the displacement of the dipole's center of mass relative to the electron, while the high-frequency one is associated with the internal structure of the dipole. At small $d/a$, the low-frequency optical phonon softens with $\omega_d \propto \sqrt{\partial^2_r V(r)\eval_{r=a}/m} \propto \sqrt{ d^2/m a^5}$ (where $V(r) \propto d^2/r^3$ is the electron-dipole potential), hence the zero-point displacement increases and is given by:
\begin{eqnarray}
\xi_d \sim \sqrt{\frac{\hbar}{m\omega_d}} \propto a (\frac{a a_B}{d^2})^{1/4}. \label{criterion}
\end{eqnarray} 
In the classical limit $a_B/a \rightarrow 0, a_B/d \rightarrow 0$ with $a/d$ fixed, $\xi_d \ll a$ ensures the stability of the classical composite crystal against quantum fluctuation.    

For small $d/a$, Eq.\eqref{criterion} implies that quantum melting of the dipole (=interlayer exciton) sublattice in the composite crystal occurs first when $a_B$ reaches the order of $d^2/a$. Meanwhile, the stronger electron-electron repulsion renders the electron lattice stable against quantum fluctuations until $a_B$ reaches the order of the inter-particle distance $a$, as for the canonical Wigner crystal. This naturally raises the question of what the ground state is at intermediate $a_B$ between $d^2/a$ and $a$, where both quantum effects and electrostatic interactions play crucial roles. 

\begin{figure}[t!]
  \centering
  \includegraphics[width=0.4\textwidth]{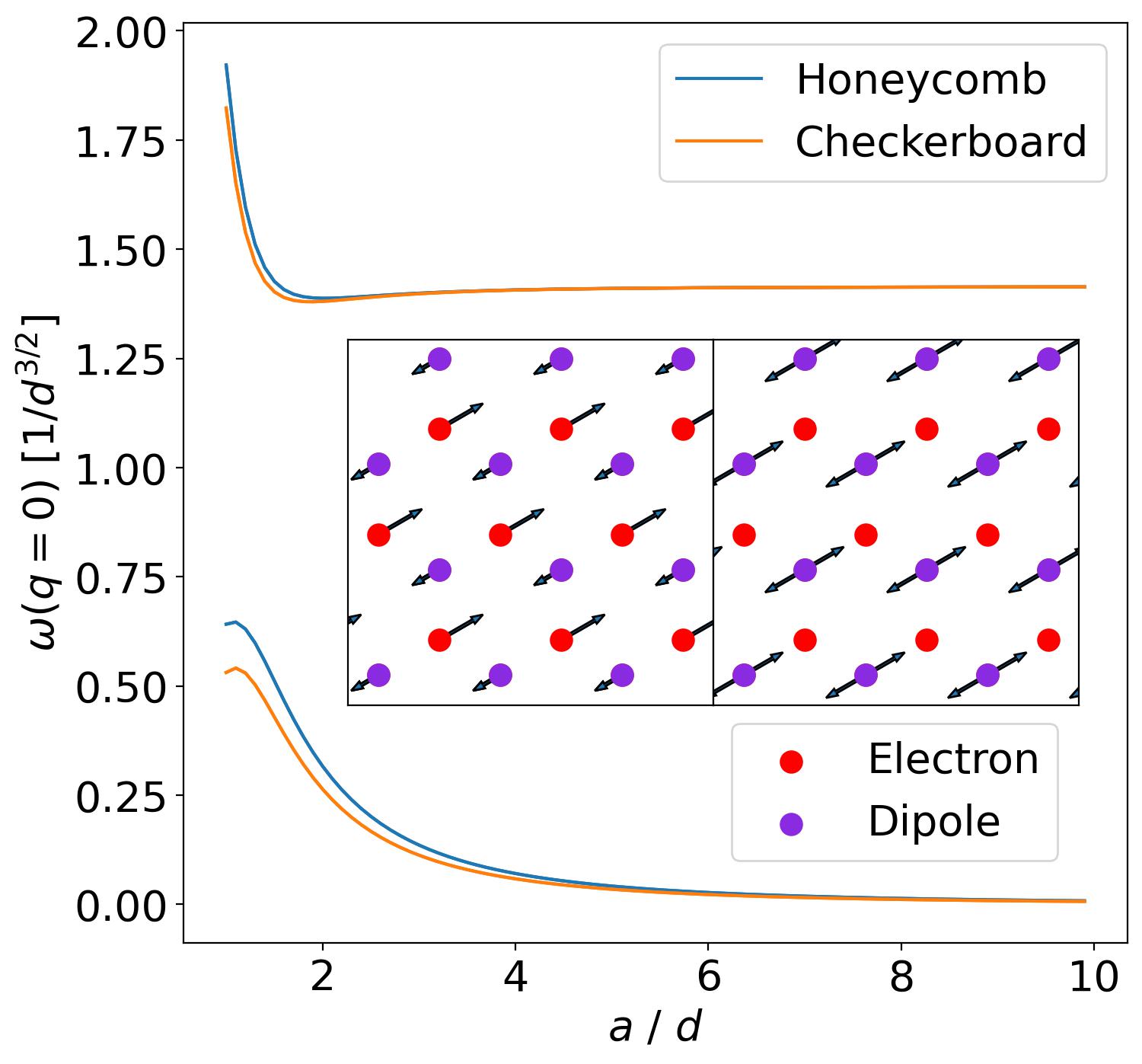}
  \caption{Soft and hard optical phonon frequencies at the center of the Brillouin zone as function of $a / d$ for the classical composite crystals. The low-frequency branches correspond to the displacement of the dipole's center of mass relative to the electron, while the high-frequency branches are associated with oscillations in the dipole's internal structure.}
  \label{fig:phonon}
\end{figure}

\textit{Dilute Quantum Regime ---} Of particular interest is the low-density limit defined by $a\rightarrow \infty$ while keeping $d$ and $a_B$ fixed, especially if $d$ and $a_B$ are of the same order. With $a \rightarrow \infty$, $a_B$ necessarily lies between $d^2/a$ and $a$, forcing partial melting of the composite crystal due to the zero-point motion of the excitons' center of mass.

The ground state in this low-density limit depends crucially on the ratio of the Bohr radius $a_B$ and layer distance $d$. For $d < d_c \approx 15.38 a_B$ (assuming equal electron and hole masses), it is known from quantum Monte Carlo  calculations that the lowest-energy state of two electrons and one hole is a trion bound state with the two electrons in a spin-singlet \cite{Witham2018}. The spin-triplet trion is absent for $m_e=m_h$ and has a smaller binding energy when it is present for large mass ratio $m_e/m_h$ \cite{Witham2018, biexciton_indistinguishable}. The trion's binding energy is on the order of $E_h$ at $d=0$ and decreases as $d$ approaches the critical $d_c$ at which the trion unbinds. At a given $d<d_c$ and in the dilute limit $a\rightarrow \infty$, the hierarchy of energy scales is necessarily as follows: $E_\text{trion} \gg E_\text{Coulomb} \sim 1/a \gg E_\text{kinetic} \sim 1/a^2$.

The physically realistic parameters for the TMD bilayer $\text{WSe}_2$/$\text{MoSe}_2$ are electron effective mass $m_e(\text{MoSe}_2) = 0.8 m_0$ ($m_0$ is the bare mass), hole effective mass $m_h(\text{WSe}_2) = 0.4 m_0$, dielectric constant $\epsilon = 4.7\epsilon_0$, and interlayer spacing $d = 3 \text{ nm}$ \cite{MoSe2_me, WSe2_me_1, WSe2_me_2, hBN_epsilon}. With these numbers, the trion binding energy obtained from quantum Monte Carlo studies \cite{Witham2018} is on the order of $15 \text{ K}$. Our Hartree-Fock calculations confirm that the ground state with these parameters and a realistic trion density of $n_\text{t} = n_\text{h} = 1.29\cdot 10^{12} \text{ cm}^{-2}$ (which corresponds to a trion density parameter $r_{s, \text{trion}} = 40$ and hole density parameter $r_{s, \text{hole}} = 8$) is an insulating quantum crystal \cite{supplement}.

We note that although Hartree-Fock qualitatively captures many phases, including the trion Wigner crystal, it is generally biased towards symmetry-breaking phases and may not produce accurate phase boundaries. For our 2 : 1 bilayer, it is important to combine Hartree-Fock with more advanced numerical methods to accurately determine the complete phase diagram. We have recently applied variational neural network wavefunctions to the balanced electron-hole bilayer and are planning a comprehensive study of the 2 : 1 bilayer with this technique in the future \cite{GemiNet}.

\begin{figure}[t!]
  \centering
  \includegraphics[width=0.45\textwidth]{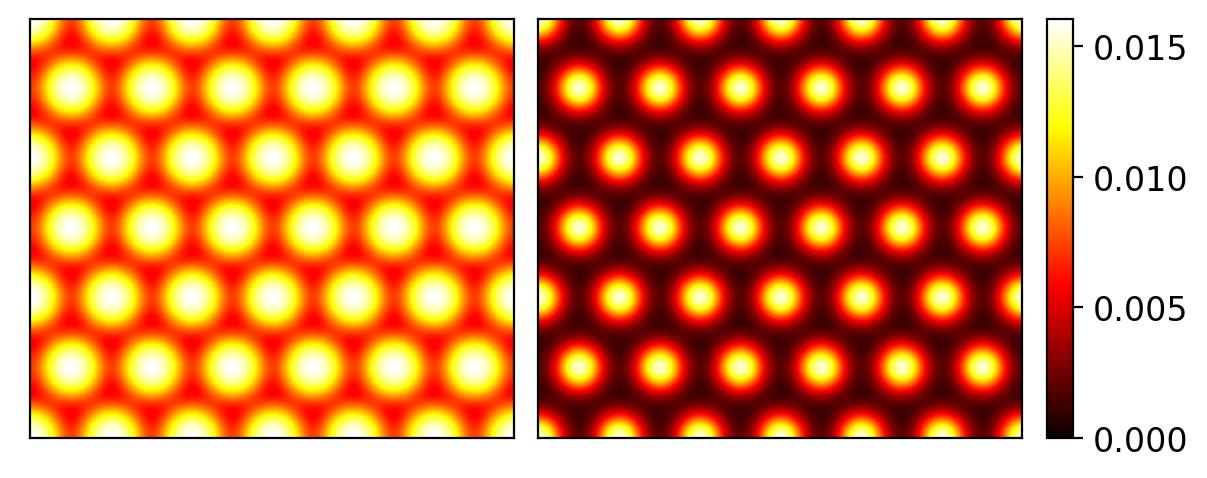}
  \caption{The electron (hole) density is shown in the left (right) panel. The coloring scheme is consistent for both densities and is indicated by the colorbar in units of $1/a_{B, h}^2$ for $a_{B, h} = \frac{4\pi\epsilon\hbar^2}{e^2 m_h}$. The calculation parameters are $m_e = 0.8m_0$, $m_h = 0.4m_0$, $d=4.83 a_{B,h}$, hole $r_s = 8$, $n_h = 36$, and equal electron spin-populations.}
  \label{fig:singlet crystal}
\end{figure}

Notably, trions can crystallize at significantly higher densities than electrons or holes. Due to the trion's large binding energy $\sim 15\text{K}$ and small spatial extent $\sim d$, in the trion Wigner crystal phase each trion behaves as charge-$e$ fermion similar to an electron in an electron Wigner crystal. But because the trion mass can be several times the electron mass, the trion's Bohr radius is several times smaller, so the same critical density parameter $r_s=30\sim40$ corresponds to a much higher electron / hole density. For example, the $2e-1h$ trion in $\text{WSe}_2$/$\text{MoSe}_2$ has total mass $2m_0$, allowing density parameter $r_s = 40$ to be reached at hole density $1.29 \cdot 10^{12} \text{ cm}^{-2}$. At these high densities, the effect of disorder and finite temperature is less severe, because all energy scales are larger. We also note that the possibility of trion crystallization in Van der Waals heterostructures has been considered recently \cite{Bondarev_2021, Bondarev_2022}.

Another distinctive feature of our trion Wigner crystal is that the electron layer has zero total spin and a large spin gap on the order of $E_h$ due to the trion binding energy, whereas the spins of localized holes interact with each other through a weak exchange interaction which vanishes in the low-density limit. Therefore, a small magnetic field can fully polarize the holes while leaving the spin-singlet electrons intact. This sharp contrast between the spin response of electron and hole layers is an indication of spin-singlet trion formation. As the density increases, the trion Wigner crystal could melt into a trion Fermi liquid before the trions dissociate. Such a metallic phase could be differentiated from the insulating trion Wigner crystal using transport measurements.

Finally, we discuss the regime $d>d_c$. Here, the trion is unstable at the three-particle level and unbinds into an electron and exciton. However, as shown earlier, the exciton sublattice in the composite crystal is necessarily unstable against quantum melting in the low-density limit. 
Consequently, we propose that for $d>d_c$ and sufficiently low densities, the ground state is an exciton superfluid permeating through an electron crystal. This state is remarkable as it simultaneously exhibits crystallization and superfluidity. For this reason, we call it an ``electron-exciton supersolid'', a quantum electron solid in which interstitial excitons Bose condense. Experimentally, this phase could be detected by Coulomb drag measurements. We also note that supersolid phases of excitons have been studied in density-balanced electron-hole bilayers \cite{peeters, sarma1, sarma2, sarma3}.

We emphasize that our predicted phases are robust against perturbations to the model Hamiltonian in Eq. \ref{eq:model_hamiltonian}. For example, deviations from the $1 / r$ potential at short distances are expected, which modifies the exciton and trion binding energies \cite{Semina_2019}. This shifts the phase boundaries (for example between the electron-exciton supersolid and trion crystal), but does not change the many-body phases that we established in the dilute limit. We also note that coupling to phonons may lead to polaron formation \cite{interfacial_polaron}, which would enhance the already high effective mass and make our phases even more robust.

\begin{figure}[t!]
  \centering
  \includegraphics[width=0.30\textwidth]{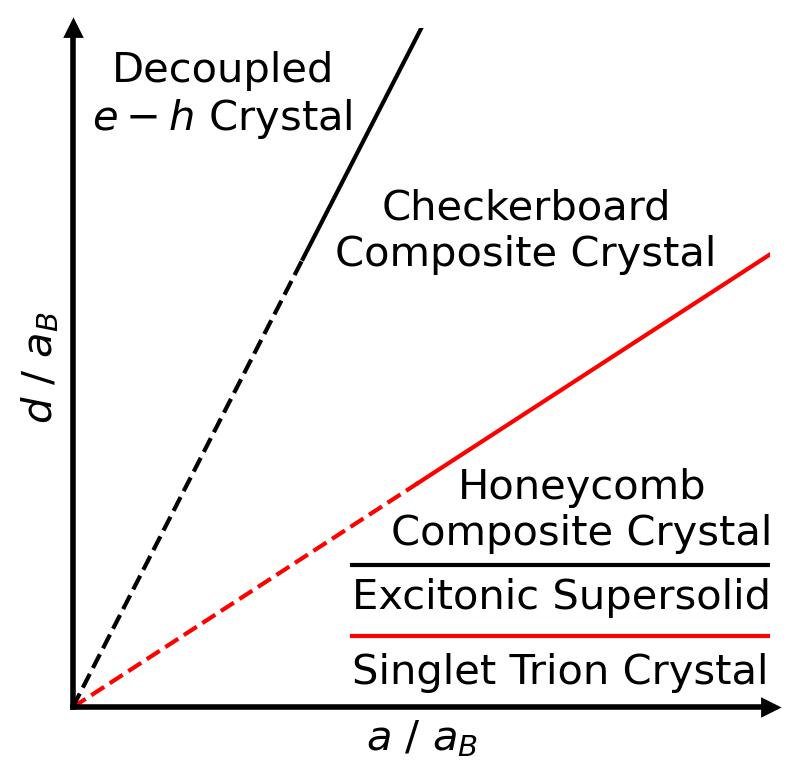}
  \caption{Schematic phase diagram for the electron-hole bilayer at commensurate densities $n_e/n_h = 2$, with analytically known boundaries marked in red and limiting phases labelled.}
  \label{fig:schematic phase diagram}
\end{figure}

\textit{Conclusion---} Our work introduces the electron-hole bilayer at commensurate 2 : 1 density ratio as a platform for realizing novel strong-coupling phases. We identify two dimensionless ratios that govern the phase diagram: the ratio of the average interparticle spacing to the Bohr radius $a / a_B$, and the ratio of the interlayer spacing to the Bohr radius $d / a_B$. By a combination of theoretical analysis and numerical calculation, we find a number of ordered phases, shown schematically in Fig. \ref{fig:schematic phase diagram}. In the classical regime $a / a_B \gg 1, d / a_B \gg 1$, we identify three crystalline phases: decoupled crystals in each layer for small $a / d$, a checkerboard ``composite crystal'' comprised of electrons and bound electron-hole dipoles for intermediate $a / d$, and a honeycomb composite crystal for large $a / d$. Focusing on the dilute regime $a \gg d, a_B$, as $d$ is decreased, we propose that large zero-point-fluctuations of the excitons partially melt the composite crystal into an ``electron-exciton supersolid'', where the unbound electrons remain crystalline but the excitons condense into a superfluid. As $d$ is decreased further to the order of the Bohr radius $a_B$, we show that the second-order Stark effect mediates an effective attractive force between electrons and excitons. For sufficiently small $d$, this attraction binds excitons to electrons to form spin-singlet trions, which then subsequently crystallize into a trion crystal. Finally, we suggest that a combination of optical spectroscopy, transport, and Coulomb drag measurements may be used to experimentally observe our proposed phases.

Notably, the trion crystal is stable throughout the low charge density regime  with a lattice constant varying continuously with the charge density $n=n_e- n_h$, provided that the electron-to-hole density ratio is maintained at the commensurate value $n_e/n_h=2$.  Thus, the trion crystal is charge-compressible with $\frac{\partial n}{\partial \mu} \neq 0$ with $\mu$ the charge chemical potential, but has an energy gap to adding excitons, i.e.,  exciton-incompressible. This should be contrasted with the excitonic insulator at charge neutrality $n=0$ which is charge-incompressible and exciton-compressible. Remarkably, recent capacitance and optical experiments on WSe$_2$/MoSe$_2$ have shown that the charge and exciton compressibility can be measured independently by varying the top and bottom gate voltages concurrently with $\Delta V_B= \pm \Delta V_T$. Moreover, the formation of trion is accompanied by a reduction of spin susceptibility in the electron (majority carrier) layer, which can be detected by magnetic circular dichroism. Finally, it is noted that the binding energy of trions can be further increased by applying an magnetic field, which we leave  to future study. We hope our theoretical work stimulates experimental study of TMD electron-hole bilayers at commensurate electron-hole density ratio.   

Note: Very recently, evidence of spin-singlet trions have been observed in optical reflectance measurements and transport experiments in $\text{WSe}_2$/$\text{MoSe}_2$ \cite{Feng_Wang_trion, KFM_trion}.

\textit{Acknowledgements---} We are grateful to Trithep Devakul and Aidan Reddy for helpful discussions. We thank Kin Fai Mak, Jie Shan and Feng Wang for their interest and feedback. This work was supported by the Simons Investigator Award from the Simons Foundation. DDD was supported by the Undergraduate Research Opportunities Program at MIT. The authors acknowledge the MIT SuperCloud and Lincoln Laboratory Supercomputing Center for providing high performance computing.

\bibliography{bibliography.bib}



\clearpage

\onecolumngrid
\begin{center}
\textbf{\large Strong-coupling phases of trions and excitons in electron-hole bilayers at commensurate densities: Supplementary Material}\\[0.5cm]
David D. Dai and Liang Fu\\
{\itshape{\small Department of Physics, Massachusetts Institute of Technology, Cambridge MA 02139, USA}}\\
\end{center}
\twocolumngrid

\setcounter{page}{1}
\setcounter{equation}{0}
\setcounter{figure}{0}
\setcounter{table}{0}
\setcounter{section}{0}

\renewcommand{\v}[1]{{\mathbf{\boldsymbol{#1}}}}
\renewcommand{\theequation}{S\arabic{equation}}
\renewcommand{\thefigure}{S\arabic{figure}}
\renewcommand{\thetable}{S\arabic{table}}

\tableofcontents

\section{Ewald Summation}
We use Ewald summation to efficiently calculate the classical electrostatic energy and determine the classical phase diagram. Although the principle is the same as in three dimensions, there are some additional subtleties in our case, which we discuss here \cite{bilayer_ewald}. Without Ewald summation, the $v(r) = 1 / r$ potential converges very slowly or not at all in real space, and likewise for its Fourier transform $v(k) = \int \text{d}^2\textbf{r} [v(r)e^{-i\textbf{k}\cdot\textbf{r}}] =  2\pi/k$ in reciprocal space. Consider the interlayer potential $v(r) = 1 / \sqrt{d^2 + r^2}$. Its Fourier transform $v(k) = 2\pi e^{-kd}/k $ converges for finite $d$, but convergence becomes very slow for small $d$, rending direct summation in reciprocal space inefficient. To solve this problem, we split the potential into a short-range part and a long-range part as shown below:
\begin{equation}\label{short and long range}
\begin{split}
    v_\text{short}(r) &= \frac{1-\erf{(\frac{\sqrt{d^2 + r^2}}{r_0})}}{\sqrt{d^2 + r^2}},\\
    v_\text{long}(r) &= \frac{\erf{(\frac{\sqrt{d^2 + r^2}}{r_0})}}{\sqrt{d^2 + r^2}},\\
\end{split}
\end{equation}
where $r_0$ is a cutoff parameter that we may choose, and $r$ always denotes the in-plane distance. It is very important that the argument of the error function involves $\sqrt{d^2 + r^2}$ instead of just $r$, because otherwise the long-range potential would have ``fine-detail'' on the scale of $d$ as $r$ approaches $0$, rending its Fourier transform slowly convergent. The short-range potential can be summed directly in real-space without any additional trouble by summing over all interactions between a unit cell and its neighbors, where we choose the neighborhood's size based on the value of $r_0$ relative to the unit cell dimensions and our desired accuracy.

\begin{figure}[t!]
  \centering
  \includegraphics[width=0.45\textwidth]{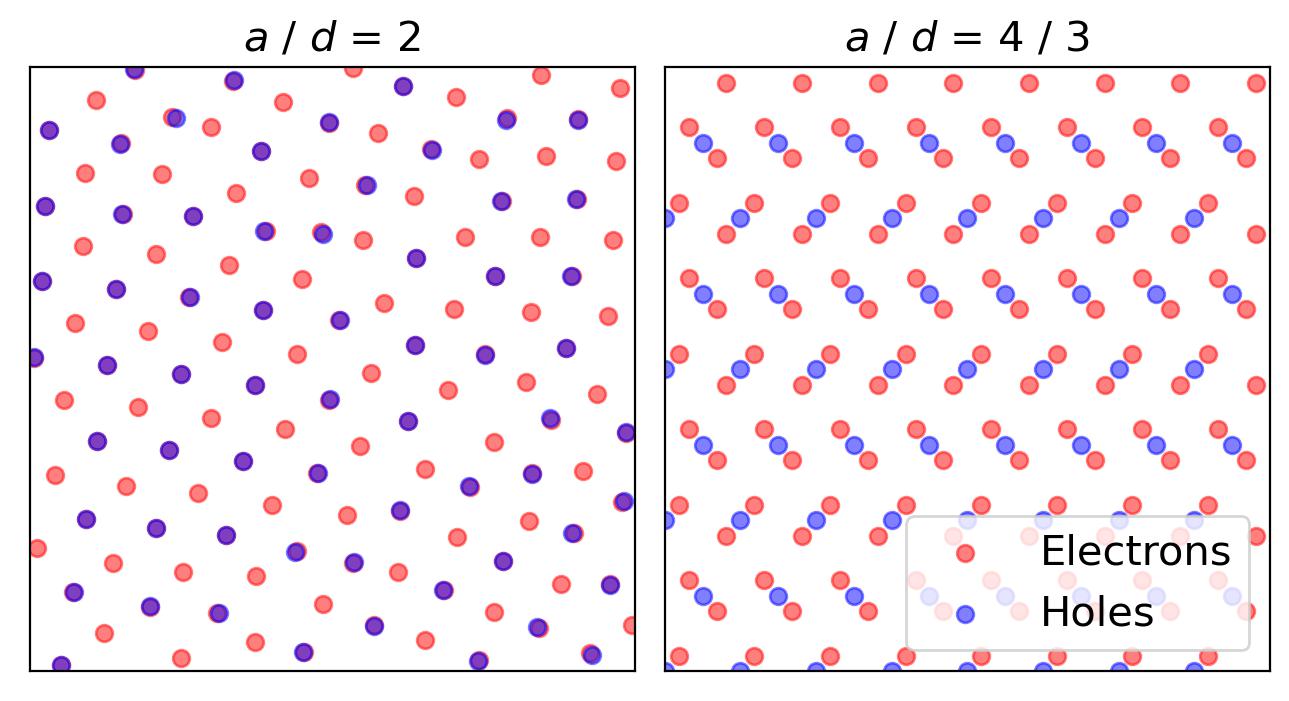}
  \caption{A local minima found by classical energy minimization with $a / d = 2$ and $64$ holes is shown on the left. Because $a / d = 2$ is denser than the crossover $a / d = 5.42$, the local minima is comprised predominantly of the checkerboard composite crystal, although there are various defects present. On the right, we show a classical ``trion crystal'' with balanced forces, where there is clear grouping into linear $e-h-e$ trions. Although stable against certain directions, such as a uniform rotation of all trions, this configuration is unstable against arbitrarily small symmetry-breaking perturbations.}
  \label{fig:additional classical}
\end{figure} 

Unlike the standard three-dimensional case, the Fourier transform of the long-range potential does not have a nice closed-form expression (Mathematica was unable to compute the Fourier transform analytically), so it is easiest to approximate it numerically as the difference between the original potential's Fourier transform and the short-range potential's Fourier transform:
\begin{equation}\label{long range FT}
\begin{split}
    v_\text{long}(k) &= v(k) - v_\text{short}(k),\\
    &= \frac{2\pi}{k}e^{-kd} - v_\text{short}(k).\\
\end{split}
\end{equation}
The short-range potential's Fourier transform can be expressed as:
\begin{equation}
\begin{split}
    v_\text{short}(k) &= \int_0^\infty r \text{d}r\bigg[\frac{1-\erf{(\frac{\sqrt{d^2 + r^2}}{r_0})}}{\sqrt{d^2 + r^2}} \int_0^{2\pi} \text{d}\theta \big[ e^{ikr\cos{\theta}} \big] \bigg],\\
    &=\int_0^\infty \text{d}r\bigg[\frac{1-\erf{(\frac{\sqrt{d^2 + r^2}}{r_0})}}{\sqrt{d^2 + r^2}} 2\pi r J_0(kr) \bigg],\\
\end{split}
\end{equation}
where we have used the integral representation of the Bessel functions of the first kind. Because the error function suppresses the integrand for large $r$, the upper limit can be replaced with $6 r_0$ ($6$ was chosen to make the error from cutting off the integral negligible) and the resulting proper integral evaluated numerically. To subtract the contribution of the uniform background charge from the short-range sum, one more integral is needed:
\begin{equation}\label{background integral}
\begin{split}
    V_\text{back} &= \int \text{d}^2\textbf{r}\bigg[v_\text{short}(r)\bigg],\\
    &= \int_0^\infty \text{d}r\bigg[2\pi r\frac{1-\erf{(\frac{\sqrt{d^2 + r^2}}{r_0})}}{\sqrt{d^2 + r^2}}\bigg],\\
    &= 2 \pi \bigg[\frac{r_0 e^{- \frac{d^2}{r_0^2}}}{\sqrt{\pi}} - d \bigg(1 - \erf{\frac{d}{r_0}}\bigg)\bigg],
\end{split}
\end{equation}
where the integral has been evaluated analytically using a computer algebra system.

\begin{figure}[t!]
  \centering
  \includegraphics[width=0.35\textwidth]{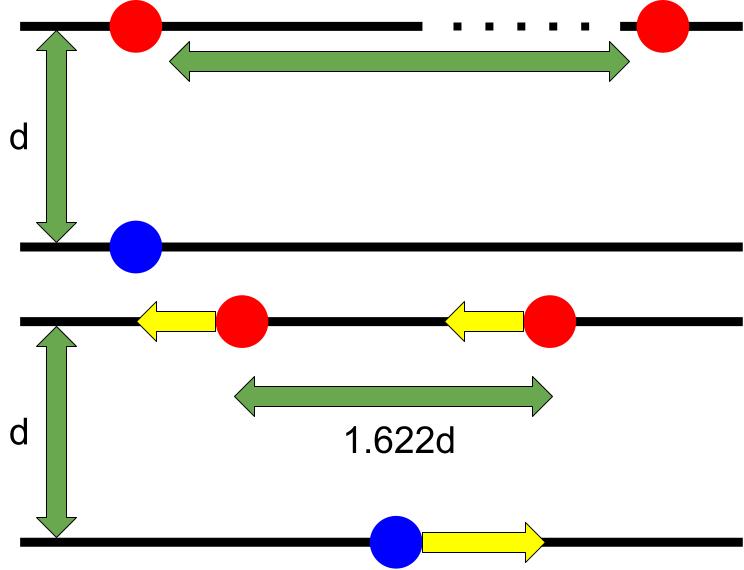}
  \caption{Arrangement of electrons (red) and holes (blue) in the exciton (top) and trion (bottom) configurations. The exciton is the unique ground state of the system with energy $-1/d$ compared to the trion's $-0.937/d$. Additionally, the trion has an imaginary frequency normal mode shown by the yellow arrows, making it a saddle point of the energy instead of a local minima.}
  \label{fig:classical trion}
\end{figure} 

The cutoff parameter $r_0$ was chosen to always be one-quarter of the smallest unit cell dimension, and both the real-space and reciprocal-space sums were converged to around eight decimal places. For our exploratory calculations, we scanned through various $a / d$ and placed $128$ electrons and $64$ holes in a large supercell to avoid artificially imposing any symmetry on the result. For each value of $a / d$, we ran several calculations with different random initializations to ensure that we discovered all possible phases. An example from one of these calculations is shown in Fig. \ref{fig:additional classical}. Once the honeycomb and checkerboard composite crystals were discovered, we used the defect-free geometries to determine the structures' precise energies.

One may wonder whether a lattice of classical trions may be stable. As with the single trion, we were able to find force-balanced configurations that are periodic arrays of trions, as shown in Fig. \ref{fig:additional classical}. However, even the slightest symmetry-breaking perturbation destroys the classical trion crystal, and the system relaxes into a composite crystal of electrons and dipoles, indicating that the classical trion crystal is a saddle point and not a local minima of the classical energy. Indeed, the trion crystal relaxes to the composite crystal through the same instability that the single trion has, where one of the electrons approaches the hole and binds with it to form a dipole while the remaining electron leaves (shown in Fig. \ref{fig:classical trion}). These results are consistent with the total absence of trion configurations in our randomly initialized calculations.

\section{Dilute-Limit Reduction of Trion to Point Particle}

We explicitly prove that in the dilute limit, where the trion's spatial extent is negligible compared to the inter-trion distance, the trion Wigner crystal behaves identically to a crystal of equal-mass point particles (and more generally, that any composite quantum particle such as a molecule is equivalent to an equal-mass point particle once its internal motion is integrated out). The Hamiltonian for a composite quantum particle such as the trion is:
\begin{equation}
    H = \sum_{i=1}^{N}\frac{1}{2m_i}\nabla_i^2 + V(\{\v r_i - \v r_j\}),
\end{equation}
where there are $N$ constituent particles, $m_i$ is the mass of the $i$th constituent, $\nabla_i$ is the gradient with respect to the $i$th constituent, and the internal potential $V$ only depends on the separations between the particles. For convenience, define the total mass $M = \sum_i m_i$.

Suppose that we transform to $N-1$ relative coordinates $\v r'_{1, \cdots, N-1}$ and $1$ center-of-mass (COM) coordinate $\v r'_N$ using linear transformation $\v r_i' = \Lambda_{ij} \v r_j$. The potential energy now has some arbitrary dependence on the relative coordinates $V(\{\v r'_{1,\cdots,N-1}\})$ but does not depend on the COM coordinate $\v r'_N$ due to translation invariance. Meanwhile, because the gradient transforms as $\nabla_i = \nabla_j' \Lambda_{ji}$, the kinetic energy in the transformed coordinates is:
\begin{equation}
    T = \sum_{i=1}^{N}\frac{1}{2m_i}\nabla_i^2 = \frac{1}{2} \nabla_j' \nabla_k' \frac{F_{ji} F_{ki}}{m_i},
\end{equation}
where for brevity we may define diagonal matrix $B_{ij} = \delta_{ij}/m_i$ and $\v S = \v F \v B \v F^T$to yield:
\begin{equation}
    T = \frac{1}{2}\nabla_j' \nabla_k' S_{jk}.
\end{equation}
We aim to show that the COM coordinate only appears in $T$ as $\nabla'^2_N/2M$ and does not mix with the relative coordinates, or in other words that $S_{iN} = \delta_{iN}/M$.

For $\v{r}'_{1,\cdots,N-1}$ to be relative coordinates, they must be invariant under the translation $\v r_i \rightarrow \v r_i + \v a$ applied to all $i$. Additionally, because $\v r'_N$ is the COM, the last row of $\v F$ is $F_{N,j} = m_j / M$. Then the sum of $\v F$'s rows satisfies:
\begin{equation}
    \sum_j F_{ij} = \delta_{iN}.
\end{equation}
Using this, we have:
\begin{equation}
    [\v B \v F^T]_{iN} = \frac{1}{m_i} F_{Ni} = \frac{1}{M},
\end{equation}
so $S_{iN} = \sum_j F_{ij} [\v B \v F^T]_{jN} = \sum_j F_{ij}/M = \delta_{i,N}/ M$, as desired. Suppose that the composite particle sits in an external potential that only varies significantly at the large length scale $a$, and let $d$ be the spatial extent of the composite particle. Then at lowest order in $d / a$, coupling between the internal degrees of freedom and external potential is negligible, and the composite particle behaves like a point particle of mass $M$ located at its COM.

\section{Hartree-Fock Derivation}

To study the quantum phases of the electron-hole bilayer, we introduce a mean-field approximation that generalizes the Slater determinant of Hartree-Fock theory. Consider a system of fermions of various species $1 \dots f$, where fermions of the same species are identical. For example, species could distinguish between different spin polarizations, between electrons and holes, or between valley degrees of freedom. A wavefunction describing this system should be antisymmetric under the exchange of two same-species fermions, but this condition does not apply to the exchange of different-species fermions, as such an exchange does not really make sense. A simple ansatz for this system is a product of Slater determinants, one for each species. Letting $\mathbf{r}_i^\alpha$ denote the position of the $i$th fermion of species $\alpha$ (throughout this paper, we use the Greek letters starting with $\alpha$ and $\beta$ to index species) and $N_\alpha$ denote the number of fermions of species $\alpha$, we can write this ansatz as:
\begin{equation}
\begin{aligned}
 &\psi(\mathbf{r}^1_{1}, \ldots, \mathbf{r}^1_{N_1}; \mathbf{r}^2_{1}, \ldots, \mathbf{r}^2_{N_2} ;\ldots; \mathbf{r}^f_{1}, \ldots, \mathbf{r}^f_{N_f}) \\
 &= 
 \prod_{\alpha} \frac{1}{\sqrt{N_\alpha!}}
 \begin{vmatrix} \phi^\alpha_1(\mathbf{r}^\alpha_1) & \phi^\alpha_2(\mathbf{r}^\alpha_1) & \cdots & \phi^\alpha_{N_\alpha}(\mathbf{r}^\alpha_1) \\
                      \phi^\alpha_1(\mathbf{r}^\alpha_2) & \phi^\alpha_2(\mathbf{r}^\alpha_2) & \cdots & \phi^\alpha_{N_\alpha}(\mathbf{r}^\alpha_2) \\
                      \vdots & \vdots & \ddots & \vdots \\
                      \phi^\alpha_1(\mathbf{r}^\alpha_{N_\alpha}) & \phi^\alpha_2(\mathbf{r}^\alpha_{N_\alpha}) & \cdots & \phi^\alpha_{N_\alpha}(\mathbf{r}^\alpha_{N_\alpha})
  \end{vmatrix},
\end{aligned}
\end{equation}
where the $\phi^\alpha_{1 \cdots N_\alpha}$'s are orthonormal single-particle orbitals (orthonormality is only enforced between orbitals belonging to the same species; there is no need for conditions on orbitals belonging to different species). By comparing to the single-determinant ansatz, we can see that our multi-determinant ansatz is indeed normalized. While it is true that such a product of Slater determinants cannot capture correlations between particles of the same or different species, this ansatz is a good place to start for a first analysis, and we expect it to be adequate for describing many strong-coupling phases that we are interested in.

To describe the most general system of $f$ interacting species, let species $\alpha$ experience a single-particle Hamiltonian $h^{\alpha}$, and let particles of species $\alpha$ and $\beta$ experience an interaction potential $v^{\alpha \beta}$. The many-body Hamiltonian is:
\begin{equation}
    H = \sum_{\alpha} \sum_{i=1}^{N_\alpha} h^{\alpha}(\mathbf{r}_{i}^{\alpha}) + 
    \frac{1}{2} \sum_{\alpha \beta}\sum_{i=1}^{N_\alpha}
    \sum_{j=1}^{N_\beta} v^{\alpha \beta}(\mathbf{r}_{i}^{\alpha} - \mathbf{r}_{j}^{\beta}).
\end{equation}
Technically, we should exclude $i=j$ in the sum when $\alpha = \beta$, but this self-interaction will be exactly cancelled by the exchange energy, so we need not worry about it. Following the same procedure used to calculate the expected energy of a single Slater determinant, we can show that the expected energy of our ansatz is:
\begin{widetext}
\begin{equation}
    E = \sum_\alpha \sum_{i=1}^{N_\alpha} \langle \phi^\alpha_i | h^\alpha | \phi^\alpha_i \rangle
    + \frac{1}{2} \sum_{\alpha \beta} \sum_{i=1}^{N_\alpha} \sum_{j=1}^{N_\beta} \bigg[\langle \phi^\alpha_i \phi^\beta_j | v^{\alpha \beta}(\mathbf{r}_1^\alpha - \mathbf{r}_2^\beta) | \phi^\alpha_i \phi^\beta_j \rangle 
    - \delta^{\alpha \beta} \langle \phi^\alpha_i \phi^\beta_j | v^{\alpha \beta}(\mathbf{r}_1^\alpha - \mathbf{r}_2^\beta) | \phi^\beta_j \phi^\alpha_i \rangle \bigg].
\end{equation}
\end{widetext}
Note that the second term is a generalization of the exchange interaction between electrons of the same spin and is only between fermions of the same species, while the first term in the two-particle sector generalizes the Coulomb interaction and is between all species. To minimize this energy, we express all orbitals as a linear combination of basis functions. Although we could in principle use a different basis set for each species, it is simpler to use a common plane-wave basis, which we will index with Greek letters starting with $\mu$ and $\nu$. Defining the coefficient matrix as $|\phi^\alpha_i\rangle = C^{\alpha}_{\mu i} |\mu\rangle$ and matrix elements $h^\alpha_{\mu\nu} = \langle \mu | h^\alpha | \nu \rangle $ and $v^{\alpha \beta}_{\mu \nu \kappa \lambda}  = \langle \mu \nu | v^{\alpha \beta} | \kappa \lambda \rangle$, we can write the Hartree-Fock energy expectation value as:
\begin{widetext}
\begin{equation}
    E = 
    \sum_{\alpha} h^\alpha_{\mu \nu} C^{\alpha*}_{\mu i} C^{\alpha}_{\nu i} 
     + \frac{1}{2} \sum_{\alpha \beta} (v^{\alpha \beta}_{\mu \kappa \nu \lambda} - \delta^{\alpha \beta} v^{\alpha \beta}_{\mu \kappa \lambda \nu}) C^{\alpha*}_{\mu i} C^{\alpha}_{\nu i} C^{\beta*}_{\kappa j} C^{\beta}_{\lambda j},
\end{equation}
\end{widetext}
where Einstein summation is implied for all indices except the species indices $\alpha$ and $\beta$. Defining the one-particle reduced density matrix as $D^{\alpha}_{\mu \nu} = C^{\alpha*}_{\mu i} C^{\alpha}_{\nu i}$, we can further simplify the energy expression into:
\begin{equation}\label{HF energy 1}
\begin{aligned}
    E = \bigg[\sum_{\alpha} h^{\alpha}_{\mu \nu}
     + \frac{1}{2} \sum_{\alpha \beta} \bigg( v^{\alpha \beta}_{\mu \kappa \nu \lambda} - \delta^{\alpha \beta} v^{\alpha \beta}_{\mu \kappa \lambda \nu} \bigg) D^{\beta}_{\kappa \lambda}\bigg]D^{\alpha}_{\mu \nu}.
\end{aligned}
\end{equation}

In what follows, we demonstrate a way of directly minimizing the Hartree-Fock energy functional using unconstrained minimization techniques instead of the standard self-consistent field (SCF) method \cite{Arias1992}. By analytically continuing the energy functional onto the space of non-orthonormal orbitals using Lowdin symmetric orthogonalization, no constraints need to be enforced and no Lagrange multipliers are necessary. Although not commonly used, direct minimization is implemented in some density functional theory codes, and it is equivalent to the usual way of solving the Hartree-Fock equations self-consistently \cite{jdftx}. We chose this approach because it is immune from instabilities such as charge sloshing and is guaranteed to converge from any starting point, which is optimal for exploratory calculations.

If the density matrix is varied by $\Delta D^{\alpha}_{\mu \nu}$, the energy changes by an amount:
\begin{equation}\label{fock}
\begin{aligned}
    &\Delta E = \sum_{\alpha} \Tr[ \mathbf{F}^{\alpha} (\Delta \mathbf{D}^{\alpha})^{\mathsf T}], \\
    &F^{\alpha}_{\mu \nu}[D] = h^{\alpha}_{\mu \nu} + \sum_\beta \bigg( v^{\alpha \beta}_{\mu \kappa \nu \lambda} - \delta^{\alpha \beta} v^{\alpha \beta}_{\mu \kappa \lambda \nu} \bigg) D^{\beta}_{\kappa \lambda}.
\end{aligned}
\end{equation}
In the top expression, bold font denotes that an object should be treated as a matrix. Ordinarily, we would minimize the energy under the orthonormality constraint using Lagrange multipliers. However, there is a way out that allows us to convert the problem into unconstrained minimization. For any arbitrary orbitals $|i\rangle$, we can perform Lowdin symmetric orthogonalization to define the orthonormal orbitals $|\Tilde{i}\rangle = | j \rangle (S^{-1/2})_{ji}$, where $S^{-1/2}$ denotes the matrix inverse-square-root of $S_{ij} = \langle i | j \rangle$. Compared to other methods of orthogonalization such as the Gram-Schmidt process, Lowdin orthogonalization manifestly treats all orbitals on equal footing, and it yields the orthonormal orbitals ``closest'' to the original non-orthonormal orbitals in the least-squares sense \cite{Mayer_2002}. Using this procedure, the density matrix corresponding to an arbitrary $C^{\alpha}_{\mu i}$ after orthogonalization is:
\begin{equation}
    D^{\alpha}_{\mu \nu} = C^{\alpha*}_{\mu j} C^{\alpha}_{\nu k} (S^{\alpha, -1})_{k j},
\end{equation}
where $\textbf{S}^{\alpha, -1}$ is the matrix inverse of $S^{\alpha}_{ij} = C^{\alpha*}_{\mu i} C^{\alpha}_{\mu j}$. By first orthogonalizing before plugging into the energy expression, the Hartree-Fock energy can be analytically continued onto the space of non-orthonormal orbitals, enabling unconstrained minimization to be used. 

Note that the matrix inverse satisfies the relation $\Delta \mathbf{S}^{-1} = - \mathbf{S}^{-1} (\Delta \mathbf{S}) \mathbf{S}^{-1}$, and the overlap matrix can expressed as $\mathbf{S} = \textbf{C}^{\dagger} \textbf{C}$. Therefore, if we vary $\mathbf{C}^{\dagger}$ by $\Delta \mathbf{C}^{\dagger}$ (we treat variables and their complex conjugates as independent here, which we will justify later), the change in the density matrix is (we have suppressed the species index for brevity so everything here refers to a single species):
\begin{equation}
\begin{aligned}
    \Delta \mathbf{D}^{\mathsf T} &= \Delta (\mathbf{C} \mathbf{S}^{-1} \mathbf{C}^{\dagger}),\\
    &=\textbf{C}\Delta(\textbf{S}^{-1})\textbf{C}^\dagger + \textbf{C}\textbf{S}^{-1}\Delta\textbf{C}^\dagger\\
    &= - \textbf{C} \mathbf{S}^{-1} (\Delta \mathbf{S}) \mathbf{S}^{-1} \textbf{C}^\dagger + \textbf{C} \textbf{S}^{-1} \Delta \textbf{C}^{\dagger},\\
    &= \textbf{C} \textbf{S}^{-1} \Delta \textbf{C}^\dagger (\textbf{I} - \textbf{C} \textbf{S}^{-1} \textbf{C}^\dagger).\\
\end{aligned}
\end{equation}
Substituting this into (\ref{fock}) yields the variation in the energy:
\begin{equation}
\begin{aligned}
    \Delta E &= \sum_{\alpha} \Tr[ \mathbf{F}^{\alpha} \textbf{C}^{\alpha} \textbf{S}^{\alpha, -1} \Delta \textbf{C}^{\alpha, \dagger} (\textbf{I} - \textbf{C}^{\alpha} \textbf{S}^{\alpha, -1} \textbf{C}^{\alpha, \dagger})],\\
    &= \sum_{\alpha} \Tr[\Delta \textbf{C}^{\alpha, \dagger} (\textbf{I} - \textbf{C}^{\alpha} \textbf{S}^{\alpha, -1} \textbf{C}^{\alpha, \dagger})\mathbf{F}^{\alpha} \textbf{C}^{\alpha} \textbf{S}^{\alpha, -1}].\\
\end{aligned}
\end{equation}
Therefore, we have the derivative of the analytically continued energy with respect to $\textbf{C}^{\dagger}$:
\begin{equation}\label{gradient}
    \frac{\partial E}{\partial C^{\alpha*}_{\mu \nu}} = [(\textbf{I} - \textbf{C}^{\alpha} \textbf{S}^{\alpha, -1} \textbf{C}^{\alpha, \dagger})\mathbf{F}^{\alpha} \textbf{C}^{\alpha} \textbf{S}^{\alpha, -1}]_{\mu\nu}.
\end{equation}
Using the expression for the Fock matrix $\textbf{F}^\alpha$, the energy expression Eq. \ref{HF energy 1} can be rewritten to
\begin{equation}\label{fock energy}
    E = 
    \frac{1}{2} \sum_{\alpha} (h^\alpha_{\mu \nu} + F^\alpha_{\mu \nu}) D^{\alpha}_{\mu \nu},
\end{equation}
allowing the computationally-expensive Fock matrix to be reused in both energy and gradient computations. 

In general, if a real-valued function $g$ depends on the complex-valued argument $z$, we have the following identities:
\begin{equation}
\begin{aligned}
    \frac{\partial g}{\partial \Re[z]} &= 2\Re[\frac{\partial{g}}{\partial z^{*}}],\\
    \frac{\partial g}{\partial \Im[z]} &= 2\Im[\frac{\partial{g}}{\partial z^{*}}],\\
\end{aligned}
\end{equation}
where derivatives with respect to the real (imaginary) part of $z$ hold the imaginary (real) part constant, and derivatives with respect to $z^{*}$ treat $z$ and $z^{*}$ as independent. For the purposes of updating variables during some optimization algorithm such as conjugate gradient, one can pretend that $2\frac{\partial g}{\partial z^{*}}$ is the derivative of $g$ with respect to $z$, as this produces the same variable update as treating the real and imaginary parts of $z$ as separate independent variables. Therefore, because we have computed both the value and derivative of the analytically continued energy, off-the-shelf minimization algorithms such as conjugate gradient can be used to solve for the Hartree-Fock orbitals. 

For our calculations, we use a large $L_x$ by $L_y$ rectangular box with periodic boundary conditions, which is equivalent to placing our system on a torus. Implementing Hartree-Fock in a large periodic box is essentially equivalent to enforcing a unit cell and sampling on a particular mesh in the first Brillouin zone, where using a larger periodic box is equivalent to using a finer mesh, all else being constant. As with using direct minimization, using a large periodic box is very helpful when exploring a new system with unknown symmetry. Our basis set is a rectangular grid of plane-waves consistent with the periodic boundary conditions: $\phi_{\mathbf{k}}(\mathbf{r}) = e^{i\mathbf{k}\cdot \mathbf{r}} / \sqrt{L_x L_y}$, $\mathbf{k} = (2\pi n_x/L_x, 2\pi n_y/L_y)$ for integer $n_x$ and $n_y$. The one-body contribution to the Hamiltonian is just the kinetic energy and is diagonal in this basis:
\begin{equation}\label{T matrix}
    h^\alpha_{\mu\nu} = \delta_{\mu\nu}\frac{1}{2m_\alpha}k_{\mu}^2,
\end{equation}
where $\mathbf{k}_\mu$ is the wavevector corresponding to the plane-wave with the flattened index $\mu$, and $m_\alpha$ is a species-dependent effective mass. For a general interaction potential $v(\mathbf{r}_1 - \mathbf{r}_2)$, the two-body matrix element is:
\begin{widetext}
\begin{equation}\label{v matrix 1}
\begin{aligned}
    \langle \mathbf{k}_1' \mathbf{k}_2' | v(\mathbf{r}_1 - \mathbf{r}_2) | \mathbf{k}_1 \mathbf{k}_2 \rangle 
    =& \frac{1}{(L_x L_y)^2} \int_{\text{box}}\text{d}^2\textbf{r}_1 \int_{\text{box}}\text{d}^2\textbf{r}_2\bigg[ v(\mathbf{r}_1 - \mathbf{r}_2) e^{- i \textbf{k}_1' \cdot \textbf{r}_1 - i \textbf{k}_2' \cdot \textbf{r}_2 + i \textbf{k}_1 \cdot \textbf{r}_1 + i \textbf{k}_2 \cdot \textbf{r}_2) } \bigg], \\
    =& \frac{1}{(L_x L_y)^2} \int_{\text{box}}\text{d}^2\textbf{r}_1 \bigg[e^{i(-\textbf{k}_1' - \textbf{k}_2' + \textbf{k}_1 + \textbf{k}_2) \cdot \textbf{r}_1}\bigg]\int_{\text{box}}\text{d}^2(\textbf{r}_2 - \textbf{r}_1)\bigg[v(\mathbf{r}_1 - \mathbf{r}_2) e^{i(-\textbf{k}_2' + \textbf{k}_2) \cdot (\textbf{r}_2 - \textbf{r}_1)}\bigg], \\
    =& \frac{\delta_{\textbf{k}_1' + \textbf{k}_2', \textbf{k}_1 + \textbf{k}_2}}{L_x L_y} \int_{\text{box}}\text{d}^2\textbf{r}\bigg[v(\textbf{r}) e^{i(-\textbf{k}_2' + \textbf{k}_2) \cdot \textbf{r}}\bigg],\\
    =& \frac{\delta_{\textbf{k}_1' + \textbf{k}_2', \textbf{k}_1 + \textbf{k}_2}}{L_x L_y} v(\textbf{k}_2' - \textbf{k}_2),
\end{aligned}
\end{equation}
\end{widetext}
where all wavevectors are consistent with the periodic boundary conditions, we have used the periodic boundary conditions to simplify the integral's limits, and $v(\textbf{q}) = \int \text{d}^2\textbf{r}[v(\textbf{r})e^{-i\textbf{k}\cdot\textbf{r}}]$ is the Fourier transform of the interaction potential. We can see that the matrix element is only nonzero between initial and final two-particle states of the same momentum and only depends on the magnitude of the momentum transfer. As mentioned in the section discussing Ewald summation, the Fourier transform of $v(\textbf{r}) = 1/\sqrt{r^2 + d^2}$ is $v(\textbf{q}) =  2\pi e^{-qd}/q$, which handles both the intralayer and interlayer interactions. The Fourier transform is divergent at $\textbf{q}=\textbf{0}$ because our interactions are long-ranged, but this is not an issue because the contribution of $v(\textbf{q}=\textbf{0})$ to the Fock matrix is exactly cancelled by the contribution of a neutralizing background charge. To see this, we explicitly calculate the contribution of $v^{\alpha\beta}(\textbf{q}=\textbf{0})$ to the Fock matrix:
\begin{equation}\label{q0}
    \Delta F^{\alpha}_{\mu \nu} = \delta_{\mu\nu}\sum_\beta v^{\alpha\beta}(\textbf{q}=\textbf{0})\frac{D^\beta_{\kappa\kappa}}{L_x L_y} - v^{\alpha\alpha}(\mathbf{q}=\mathbf{0}) \frac{D^\alpha_{\nu\mu}}{L_x L_y}.
\end{equation}
The potential contributed by a uniform background charge $\rho^\beta$ of species $\beta$ is simply $\int\text{d}^2\textbf{r}[v^{\alpha\beta}(\textbf{r})\rho^\beta] = v^{\alpha\beta}(\textbf{q}=\textbf{0})\rho^\beta$. The trace of the density matrix $D^\beta_{\kappa\kappa}$ is exactly the number of particles of species $\beta$, and $L_x L_y$ is the system area, so the first term in Eq. \ref{q0} is exactly canceled by incorporating a neutralizing background charge for every species. The contribution to the exchange term is slightly more subtle. The total exchange matrix is
\begin{equation}\label{exchange1}
\begin{aligned}
    K^\alpha_{\mu\nu} &= v^{\alpha \alpha}_{\mu \kappa \lambda \nu} D^\alpha_{\kappa\lambda},\\
    &= \frac{1}{L_x L_y} \sum_{\textbf{q}} v^{\alpha \alpha}(\textbf{q}) D^\alpha_{\nu+q,\mu+q},
\end{aligned}
\end{equation}
where $\mu+q$ is shorthand for the plane-wave state corresponding to the wavevector $\textbf{k}_\mu + \textbf{q}$. For a finite system size, the divergent $v(\textbf{q}=\textbf{0})$ is indeed problematic at face value. However, after replacing sum with integral in the infinite system size limit, the measure $\text{d}^2\textbf{q} = 2\pi q\text{d}q$ contributes a factor of $q$ which cancels the singularity coming from the potential $v(\textbf{q}) \propto 1/q$. Cutting out the $\textbf{q}=\textbf{0}$ singularity simply results in a small finite-size error $\propto 1/L_x L_y$ in the exchange matrix. Therefore, it is completely safe simply to set $v(\textbf{q}=\textbf{0})=0$, which both removes the singularity and implements the neutralizing background charge.

For our calculations, we used a periodic rectangle with a height-to-width ratio of $\sqrt{3}/2$ to allow a triangular lattice to fit cleanly. We also performed randomly initialized runs with differently-shaped boxes and different numbers of particles, but the results were qualitatively the same. Our basis set was a $45$ by $39$ grid of plane waves consistent with the periodic boundary conditions. We choose the specific numbers $45$ and $39$ because $39/45$ and $\sqrt{3}/2$ differ only by $0.07\%$, so the resolution in each direction is essentially the same. We wrote all of our code in Python and minimized the Hartree-Fock energy using SciPy's conjugate gradient with default settings (halt optimization when maximum component of gradient is less than $10^{-5}$). All calculations were done with $72$ electrons and $36$ holes.

\begin{figure}[t!]
  \centering
  \includegraphics[width=0.45\textwidth]{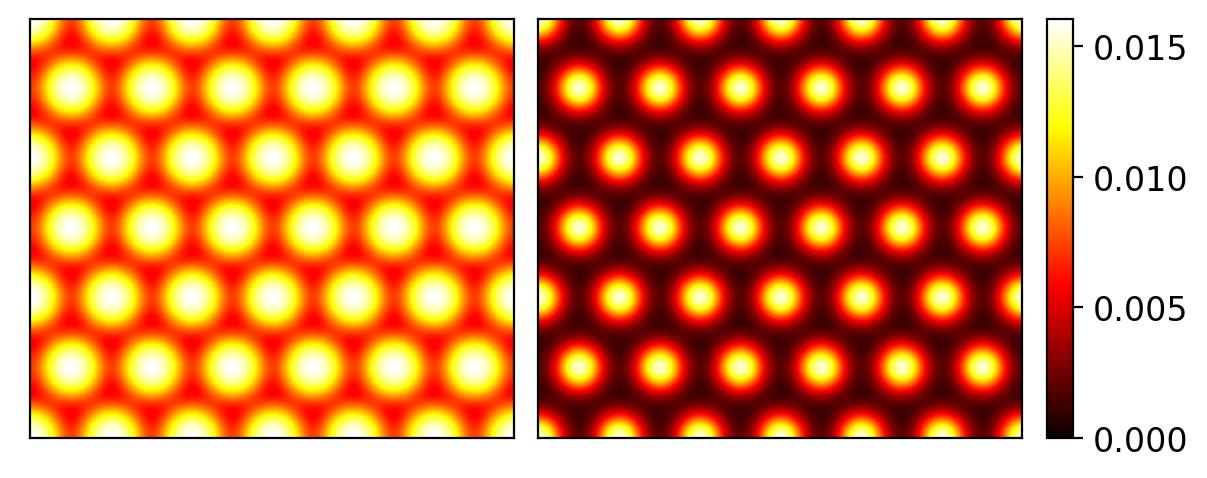}
  \caption{The total (spin $\uparrow$ plus $\downarrow$) electron density is shown in the left panel, and the hole density is shown in the right panel; the coloring scheme is consistent for both densities and is shown in the colorbar in units of $1/a_{B, h}^2$ for $a_{B, h} = \frac{4\pi\epsilon\hbar^2}{e^2 m_h}$. The calculation parameters are $m_e = 0.8m_0$, $m_h = 0.4m_0$, $d=4.83 a_{B,h}$, hole $r_s = 8$, $n_h = 36$, and equal electron spin-populations.}
\end{figure}

\section{Main Hartree-Fock Results}

\textit{Hartree-Fock Calculations---} Guided by our theoretical results in the low-density regime, we perform further numerical study to strengthen our prediction of trion Wigner crystals by applying the Hartree-Fock (HF) method to $N_e = 72$ electrons and $N_h = 36$ holes in a periodic rectangle. Our system consists of four fermion species including spin $s\in\{\uparrow, \downarrow\}$ and electron/hole degrees of freedom, and our ansatz for the HF ground state is a product over fermion species of Slater determinants. Because the ground state in the dilute limit must be a crystal of \textit{spin-singlet} trions, we perform restricted HF by forcing the electron $s=\uparrow$ and $s=\downarrow$ orbitals to be the same (no other conditions are enforced). As discussed previously, the coupling between hole spins is weak and plays an unimportant role, so we assume that the holes are fully spin-polarized, reducing the effective number of species to three.

Instead of the usual self-consistent field method, we directly minimize an analytic continuation of the Hartree-Fock energy functional onto the space of non-orthonormal orbitals \cite{Arias1992}. For any set of non-orthonormal orbitals $|i\rangle$, Lowdin symmetric orthogonalization yields the orthonormal orbitals $|\Tilde{i}\rangle = | j \rangle (S^{-1/2})_{ji}$, where $\textbf{S}^{-1}$ is the matrix inverse-square-root of the overlap matrix $S_{ij} = \langle i | j \rangle$. The HF energy of the Lowdin-orthogonalized orbitals is a sensible function over the space of non-orthonormal orbitals and is easily differentiable, allowing standard unconstrained minization techniques such as conjugate gradients to be used.

\begin{figure}[t!]
  \centering
  \includegraphics[width=0.45\textwidth]{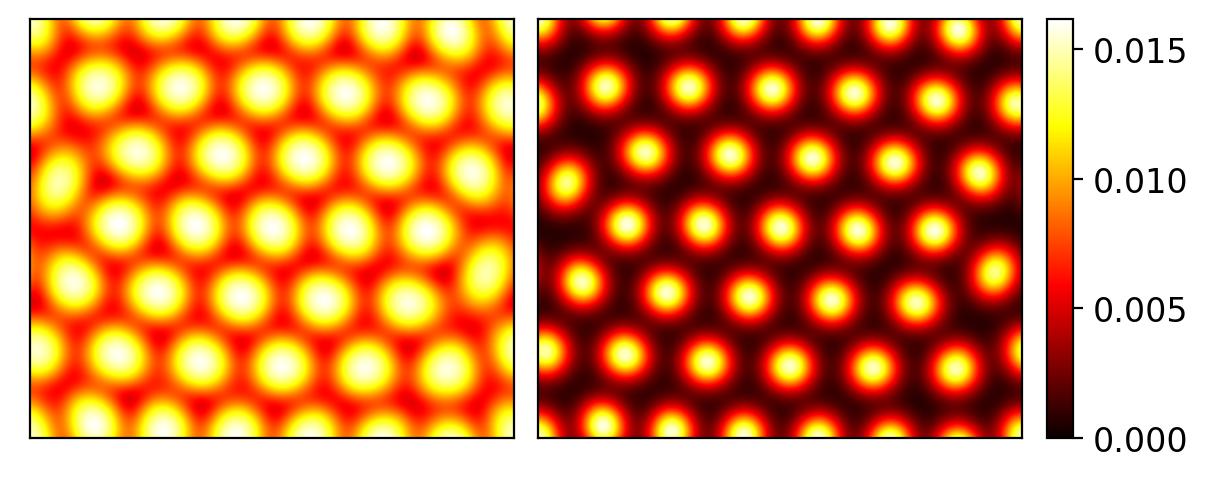}
  \caption{A local minima found by a randomly initialized RHF run at the same parameters as in the main text, with an energy per particle higher than the trion-initialized calculation by a few tenths of a percent. Both the electrons and holes have crystallized onto an approximate triangular lattice, with some defects visible on the borders of the box. }
  \label{fig: RHF_random}
\end{figure} 

Although formally equivalent to the self-consistent field method, direct minization is immune from instabilities such as charge sloshing and is guaranteed to converge from any starting point. Both of these features are convenient for exploratory calculations with unknown ground states. We supply a detailed derivation of our working equations and additional HF calculations in the supplementary materials.

The aforementioned parameters for $\text{WSe}_2$/$\text{MoSe}_2$ yield a Bohr radius for the hole layer of $a_{B, h} = 0.622 \text{ nm}$ and $d / a_{B, h} = 4.83$. We present results for trion density $n_\text{t} = n_\text{h} = 1.29\cdot 10^{12} \text{ cm}^{-2}$, which corresponds to a trion density parameter $r_{s, \text{trion}} = 40$ and hole density parameter $r_{s, \text{hole}} = 8$. 

We find that in the RHF ground state, the electrons and holes are concentrated on top of each other in a triangular lattice. The energy per particle is $-0.07652$, the electron HF spectrum has gap $0.02852$, and the hole HF spectrum has gap $0.08760$ (all energies are in terms of the hole Hartree energy). This shows that the ground state is an insulating quantum crystal of spin-singlet trions for these parameters. 

While HF provides a qualitative description of the trion crystal, it cannot capture pairing between electrons and holes in the excitonic supersolid. It is important to combine an HF study with more advanced numerical methods such as quantum Monte Carlo in order to accurately determine the parameter regions for interaction-induced phases, including the trion crystal, the composite crystals, and the putative excitonic supersolid. We leave a comprehensive numerical study of the electron-hole bilayer at commensurate $2:1$ density ratio to a future work.

\section{Additional Hartree-Fock Results}

The physically realistic parameters for the TMD bilayer $\text{WSe}_2$/$\text{MoSe}_2$ are electron effective mass $m_e(\text{MoSe}_2) = 0.8 m_0$ ($m_0$ is the bare mass), hole effective mass $m_h(\text{WSe}_2) = 0.4 m_0$, dielectric constant $\epsilon = 4.7\epsilon_0$, and interlayer spacing $d = 3 \text{ nm}$ \cite{MoSe2_me, WSe2_me_1, WSe2_me_2, hBN_epsilon}. Then the Bohr radius for the hole layer is $a_\text{B} = 0.622 \text{ nm}$ and $d / a_{B, h} = 4.83$. In the main text, we present calculations at a trion density $n_\text{trion} = n_\text{h} = 1.29\cdot 10^{12} \text{ cm}^{-2}$, which corresponds to a trion density parameter $r_{s, \text{trion}} = 40$ and hole density parameter $r_{s, \text{hole}} = 8$.

\begin{figure}[t!]
  \centering
  \includegraphics[width=0.45\textwidth]{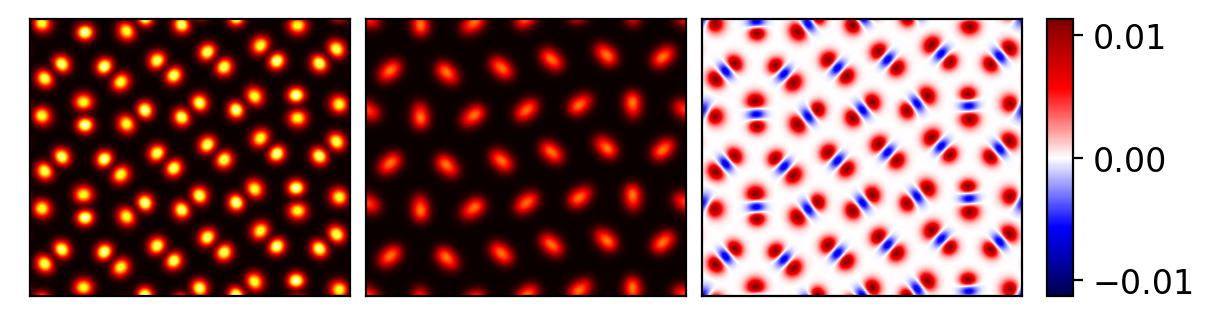}
  \caption{A local minima found by a randomly initialized HF run with electrons fully spin-polarized at hole density parameter $r_{s, h} = 16$ and all other parameters the same as in the main text. There is clear grouping into ``dumbells'' of two electrons and one hole, which are localized onto an approximate triangular lattice.}
  \label{fig: PHF_random}
\end{figure} 

We find that with random initialization, RHF converges onto a state where electrons and holes are concentrated on top of each other in an approximate triangular lattice, corresponding to a Wigner crystal of singlet trions with some defects. Due to their repulsion, the electron orbitals spread out more than the hole orbitals. Motivated by these randomly initialized calculations, we also perform RHF starting directly from singlet trions placed on a triangular lattice. This yields a lower energy than any randomly initialized calculation, indicating that the true ground state is a triangular lattice of singlet trions.

We also perform unrestricted Hartree-Fock (UHF) calculations where the electron $s=\uparrow$ and $s=\downarrow$ orbitals are not restricted to be the same, as well as spin-polarized HF calculations. We note that both unrestricted HF and HF calculations for spin-polarized electrons actually produce lower energies than RHF for the aforementioned parameters and in the dilute limit as well. However, this result is likely an artifact of the Hartree-Fock approximation because the dilute ground state should be a triangular lattice of spin-singlet trions. For the three-body problem of two electrons and one hole, HF significantly underestimates the trion binding energy or cannot even capture trion formation.  We also note that while the HF ansatz can capture the fully classical crystals composed of electrons and holes, it cannot capture pairing between electrons and holes and thus cannot describe the excitonic supersolid. Meanwhile, in the singlet trion crystal, there are strong correlations inside the trions that HF cannot describe.

To describe the trion, the effects of both kinetic energy and interaction must be taken into account. The true trion is spherically symmetric and has total angular momentum $L = 0$. Within an orbital-filling picture, this can be captured by placing all particles in colocated $s$-orbitals. This adequately treats the kinetic energy and attraction between electrons and holes, but it does not satisfy the electron-electron repulsion. Meanwhile, an orbital-filling picture in UHF can also satisfy the electron-hole attraction and electron-electron repulsion by localizing the electrons on opposite sides of the hole, but this does a poor job of minimizing the kinetic energy (by including admixture with a $p$-orbital to localize on one side of the hole) and does not respect the trion's rotational symmetry. Once the electrons are spatially separated on opposite sides of the central hole, exchange coupling makes the spin-polarized case more favorable than the spin-unpolarized case.

For most mass ratios, this spin-polarized ``dumbell trion'' with broken rotational symmetry is energetically favorable in UHF, despite it's being qualitatively incorrect. Thus, the three-body problem is not captured well by Hartree-Fock because it cannot minimize the energy while maintaining correct symmetry. By restricting the electron $s=\uparrow$ and $s=\downarrow$ orbitals to be same, we can force Hartree-Fock to yield qualitatively correct results, at the cost of having to put in a constraint by hand.

Although it is difficult to see this at the main text's parameters, at decreased density the spin-polarized HF calculations still find a crystal of trions, except that the trions are the spin-polarized type, which are shaped like a dumbell. Thus, even the spin-polarized HF calculations still confirm our general prediction of trion Wigner crystals.

\begin{figure}[t!]
  \centering
  \includegraphics[width=0.45\textwidth]{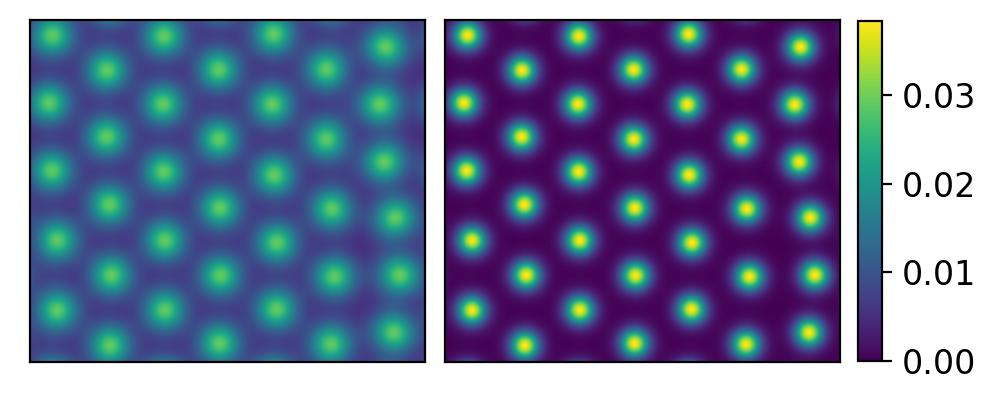}
  \caption{A local minima found by a randomly initialized UHF run at $m_e= 0.25m_0$, $m_h=m_0$, interlayer spacing $d = 0.5 a_{B,h}$, and hole density parameter $r_s = 7$. The left panel shows the total electron density and the right panel shows the density parameter (the difference between the two electron spins is visually imperceptible). The triangular lattice is nearly perfect with some defects visible on the right edge of the box .}
  \label{fig: A05R7S-R}
\end{figure} 

We also note that in the many-body problem, enforcing a net charge density increases the trion's effective region of stability relative to the electron and exciton. For an infinitely large system with $d$ larger than the critical $d_c$ at which the trion unbinds, the exciton plus electron is indeed lower in energy than the now metastable trion. However, provided that $d-d_c$ is not too large, the electron and exciton must be extremely far away from each other to actually ``break-even'' with the trion. Enforcing a net charge density effectively confines groups of two electrons and one hole into potential wells, which prevents the electron and exciton from separating to break-even distance and ``squeezes'' them back into a trion.

Additionally, we were able to find parameters for which unrestricted Hartree-Fock converges naturally to a spin-singlet: $m_e= 0.25m_0$, $m_h=m_0$, interlayer spacing $d = 0.5 a_{B,h}$, and hole density parameter $r_s = 7$. The very small electron effective mass is chosen here to impose a high cost for breaking rotation symmetry. Randomly initialized UHF calculations converge naturally onto a state where spin-up electrons, spin-down electrons, and holes are all concentrated on top of each other in an approximate triangular lattice. Additionally, the difference between the electron $s=\uparrow$ and $s=\downarrow$ densities is visually imperceptible, indicating a nearly pure electron spin-singlet. UHF calculations starting directly from singlet trions placed on a triangular lattice yielded slightly lower energies and produced gapped electron and hole spectra, indicating that the true UHF ground state at these parameters is an insulating triangular lattice of singlet trions.

\end{document}